\documentclass[aps,pra,superscriptaddress,floatfix,showpacs,10pt,twocolumn]{revtex4}
\usepackage{amsmath}
\usepackage{graphicx}
\usepackage{amsfonts}
\usepackage{amssymb}

\usepackage[T1]{fontenc}
\usepackage[latin2]{inputenc}
\usepackage{latexsym}
\usepackage{bbm}
\usepackage{CJK}
\usepackage{color}

\begin{document}
\title{Dynamics of quantum discord in the purification process }
\author{Wei Song}
 \affiliation{Department of Physics
and Electronic Engineering, Hefei Normal University, Hefei 230061,
China}
\author{Lin Chen }
\affiliation{Centre for Quantum Technologies, National University of
Singapore, 3 Science Drive 2, Singapore 117542}

\author{Ming Yang}
\affiliation{School of Physics and Material Science, Anhui
University, Hefei 230039, China }
\author{Zhuo-Liang Cao}
\affiliation{Department of Physics and Electronic Engineering, Hefei
Normal University, Hefei 230061, China}

\date{\today}


\begin{abstract}
We investigate the dynamics of quantum discord during the
purification process. In the case of Werner states, it is shown that
quantum discord is increased after a round of purification protocol.
Furthermore, quantum mutual information and classical correlation is
also increased during this process. We also give an analytic
expression for a class of higher dimensional states which have
additive quantum discord.

\end{abstract}
\maketitle

It is widedly recognized that entanglment is a key resource and
ingredient in the field of quantum information science
\cite{Neilsen:2000}. However, entanglement is not the only aspect of
quantum correlations, and it has been found that there are quantum
nonlocality without entanglement \cite{Bennett:1999,Horodecki:2005}
and quantum speedup with unentangled states
\cite{Braunstein:1999,Meyer:2000,Datta:2005}. Therefore, it is
important and interesting to investigate such quantum correlations.
In this context, several measures have been proposed to quantify the
nonclassical nature of a quantum state such as quantum discord
\cite{Henderson:2001,Ollivier:2001}, quantum deficit
\cite{Rajagopal:2002}, measurement induced disturbance
\cite{Luo:2008}. Furthermore, Modi \emph{et al.} \cite{Modi:2010}
provides an unified view of quantum and classical correlations
recently. Among these measures, quantum discord introduced by
Olliver and Zurek \cite{Zurek:2000,Ollivier:2001} is widely used to
quantify the quantumness of the correlation contained in a bipartite
quantum state. A closely related measure concerning classical
correlation was proposed by Henderson and Vedral
\cite{Henderson:2001}. Quantum discord is defined as the difference
between the total mutual information and the classical correlation
contained in a bipartite quantum state . It is more general than
entanglement in the sense that separable mixed states can have
nonzero quantum discord \cite{Dakic:2010,Chen:2010}. It has been
suggested that discord is responsible for the speed-up related to
deterministic quantum computation with one qubit(DQC1) model
\cite{Knill:1998,Datta:2008}. And quantum discord has also attracted
much attention for its connection with entanglement irreversibility
\cite{Cornelio:2011}, the entanglement monogamic relation
\cite{Fanchini:2011}, the entanglement in a
measurement\cite{Piani:2011,Streltsov:2011} and the complete
positivity \cite{Shabani:2009}. Especially, the operational
interpretations of quantum discord through the state merging
\cite{Cavalcanti:2011,Madhok:2011} and locally inaccessible
information\cite{Fanchini:2011d} have established the status of
quantum discord as another important resource for quantum
informational processing tasks besides entanglement.

On the other hand, quantum discord will degrade due to unavoidable
interaction between a quantum system and its environment. It has
been shown that quantum discord decays exponentially
\cite{Werlang:2009,Maziero:2009}, whereas the entanglement may
suffer from sudden death \cite{Yu:2009} under Markovian noises. The
dynamics of quantum discord has also been investigated in a
non-Markovian environment \cite{Wang:2010,Fanchini:2010} and has
been demonstrated experimentally under both Markovian
\cite{Xu:2010a} and non-Markovian \cite{Xu:2010b} environments. In
the field of quantum entanglement theory, to achieve a perfect
channel in quantum information tasks, entanglement purification is
performed to extract a small number of entangled pairs with a
relatively high degree of entanglement from a large number of less
entangled pairs using only local operations and classical commu-
nication (LOCC). Since entanglement is only a special kind of
quantum correlations and has different order with other quantum
correlations in general. One may ask the question whether quantum
correlation can be increased in a round of purification process. In
this work we answer this question partly in terms of quantum discord
through a concrete example. We also give an analytic expression for
a class of higher dimensional states which have additive quantum
discord.

In the following discussions we use the quantum discord to measure
the quantum correlation. As an example, we shall investigate the
dynamics of the quantum discord during the original BBPSSW
purification protocol \cite{Bennett:1996}. In the case of Werner
states,  we find that quantum discord can also be increased after
performing the original purification protocol. During one round of
purification process, the quantum discord decreases at first and
then increases when the intermediate state is transformed into the
final state under LOCC. Suppose the noisy Werner state pairs $ \rho
^{\left( 0 \right)}$
 and $ \rho ^{\left( 1 \right)}$ are given as two copies in the following
form:

\begin{eqnarray}
&\rho ^{\left( 0 \right)}&  = \rho ^{\left( 1 \right)}  = \rho  =
F\left| {\beta _{11} } \right\rangle \left\langle {\beta _{11} }
\right| \notag\\&+& {{1 - F} \over 3}\left( {\left| {\beta _{01} }
\right\rangle \left\langle {\beta _{01} } \right| + \left| {\beta
_{10} } \right\rangle \left\langle {\beta _{10} } \right| + \left|
{\beta _{00} } \right\rangle \left\langle {\beta _{00} } \right|}
\right)
\end{eqnarray}

\noindent where the four Bell states $ \beta _{ab}  \equiv {1 \over
{\sqrt 2 }}\left( {\left| {0,b} \right\rangle  + \left( { - 1}
\right)^a \left| {1,1 \oplus b} \right\rangle } \right)$. We begin
with a brief outline of the basic process of BBPSSW purification
protocol\cite{Bennett:1996}. Suppose Alice and Bob share two
identical noisy Werner state pairs $ \rho ^{\left( 0 \right)}$
 and $ \rho ^{\left( 1 \right)}$. Firstly, one of them perform $
\sigma _y $ operations on her/his particles and transform the Werner
state pairs into the form: $ F\left| {\beta _{00} } \right\rangle
\left\langle {\beta _{00} } \right| + {{1 - F} \over 3}\left(
{\left| {\beta _{10} } \right\rangle \left\langle {\beta _{10} }
\right| + \left| {\beta _{01} } \right\rangle \left\langle {\beta
_{01} } \right| + \left| {\beta _{11} } \right\rangle \left\langle
{\beta _{11} } \right|} \right)$. Next they apply a bilateral C-Not
gate for $ \rho ^{\left( 0 \right)}$
 and $ \rho ^{\left( 1 \right)}$
as the control and target qubits, respectively. Then they
bilaterally measure $ \rho ^{\left( 1 \right)}$ in the computational
basis $ \left\{ {\left| 0 \right\rangle ,\left| {\left| 1
\right\rangle } \right\rangle } \right\} $ and communicate the
measurement outcomes to each other. They keep $ \rho ^{\left( 0
\right)}$ if the measurement outcomes coincide. Otherwise, they
discard $ \rho ^{\left( 0 \right)}$. After performing the
measurement, the original mixed Werner state $ \rho ^{\left( 0
\right)}$ will be mapped into a new mixed state $\rho ' $ described
as:

\begin{eqnarray}
\rho ' &=& {{10F^2  - 2F + 1} \over {8F^2  - 4F + 5}}\left| {\beta
_{00} } \right\rangle \left\langle {\beta _{00} } \right| + {{6F -
6F^2 } \over {8F^2  - 4F + 5}}\left| {\beta _{10} } \right\rangle
\left\langle {\beta _{10} } \right| \notag\\&+& {{2F^2  - 4F + 2}
\over {8F^2 - 4F + 5}}\left| {\beta _{01} } \right\rangle
\left\langle {\beta _{01} } \right| + {{2F^2  - 4F + 2} \over {8F^2
- 4F + 5}}\left| {\beta _{11} } \right\rangle \left\langle {\beta
_{11} } \right|\notag\\
\end{eqnarray}

In fact, the mixed state $ \rho ' $ is only an intermediate state
and hence it cannot be the initial state in the second round of
BBPSSW purification. We perform $ \sigma _y $ operations firstly and
then twirl the state applying at random one of the four operations $
\left\{ {\sigma _x \otimes \sigma _x ,\sigma _y  \otimes \sigma _y
,\sigma _z \otimes \sigma _z ,I \otimes I} \right\} $
\cite{Werner:1989} locally to each party of the pair. In this way we
transform the mixed state $ \rho ' $ into the Werner state: $ \chi =
F'\left| {\beta _{11} } \right\rangle \left\langle {\beta _{11} }
\right| + {{1 - F'} \over 3}\left| {\beta _{01} } \right\rangle
\left\langle {\beta _{01} } \right| + {{1 - F'} \over 3}\left|
{\beta _{10} } \right\rangle \left\langle {\beta _{10} } \right| +
{{1 - F'} \over 3}\left| {\beta _{00} } \right\rangle \left\langle
{\beta _{00} } \right|$, with $ F' = {{10F^2  - 2F + 1} \over {8F^2
- 4F + 5}} $. The mixed state $ \chi$ is the input state in the next
round of BBPSSW purification. Because $ F' > F $ in the range of  $
{1 \over 2} < F < 1 $, using the iteration of the protocol we can
probabilistic extract a mixed state pair with a relatively high
degree of entanglement from two entangled copies using only LOCC. In
order to compare the quantum correlation during the purification
process, we compute the quantum discord of the initial mixed state $
\rho $, the intermediate state $ \rho '$ and the final state $ \chi
$, respectively.

Firstly, we recall the definitions of quantum discord. To define
quantum discord, we starts with the quantum mutual information
defined as $ \mathcal{I}\left( {\rho _{AB} } \right) = S\left( {\rho
_A } \right) + S\left( {\rho _B } \right) - S\left( {\rho _{AB} }
\right) $, where $ S\left( \rho \right) =  - Tr\left( {\rho \log _2
\rho } \right) $ is the von Neumann entropy. The quantum mutual
information is regarded as quantifying the total correlation in the
mixed state $ {\rho _{AB} } $, then the quantum discord is defined
as

\begin{eqnarray}
\mathcal{D}\left( {\rho _{AB} } \right) = \mathcal{I}\left( {\rho
_{AB} } \right) -  \mathcal{C}\left( {\rho _{AB} } \right)
\end{eqnarray}

\noindent where $ \mathcal{C}\left( {\rho _{AB} } \right) $ denote
the classical correlation of the state $ {\rho _{AB} } $ and it is
defined as $ \mathcal{C}\left( {\rho _{AB} } \right) = \mathop {\max
}\limits_{\left\{ {\Pi _k } \right\}} \left[ {S\left( {\rho _A }
\right) - S\left( {\rho _{AB} |\left\{ {\Pi _k } \right\}} \right)}
\right]$, where the maximum is taken over the set of projective
measurements $ {\left\{ {\Pi _k } \right\}} $ and $ S\left( {\rho
_{AB} |\left\{ {\Pi _k } \right\}} \right) = \sum\nolimits_k {p_k
S\left( {\rho _k } \right)} $ is the conditional entropy of system
A, with $ \rho _k  = {{Tr_B \left( {\prod _k \rho _{AB} \prod _k }
\right)} \over {p_k }} $ and $ p_k  = Tr_{AB} \left( {\rho _{AB}
\prod _k } \right) $. Here, we only consider projective measures
because Hamieh \emph{et al.} \cite{Hamieh:2004} have shown that for
a two-qubit system the projective measurement is the positive
operator-valued measure (POVM) which maximizes classical
correlation. Consider the Bell-diagonal states of the form
\cite{Lang:2010}

\begin{eqnarray}
\rho _{AB}  = {1 \over 4}\left( {I + \sum\limits_{j = 1}^3 {c_j
\sigma _j^A  \otimes \sigma _j^B } } \right) = \sum\limits_{a,b}
{\lambda _{ab} \left| {\beta _{ab} } \right\rangle } \left\langle
{\beta _{ab} } \right|
\end{eqnarray}

\noindent with eigenvalues

\begin{eqnarray}
\lambda _{ab}  = {1 \over 4}\left( {1 + \left( { - 1} \right)^a c_1
- \left( { - 1} \right)^{a + b} c_2  + \left( { - 1} \right)^b c_3 }
\right)
\end{eqnarray}

An analytical formula of quantum discord have been obtained for
Bell-diagonal states and can be expressed as \cite{Luo:2008b}:

\begin{eqnarray}
   Q &=& {1 \over 4}\left[ {\left( {1 - c_1  - c_2  - c_3 } \right)} \right.\log _2 \left( {1 - c_1  - c_2  - c_3 } \right)  \notag\\
    &+&  \left( {1 - c_1  + c_2  + c_3 } \right)\log _2 \left( {1 - c_1  + c_2  + c_3 } \right)  \notag\\
    &+& \left( {1 + c_1  - c_2  + c_3 } \right)\log _2 \left( {1 + c_1  - c_2  + c_3 } \right)  \notag\\
   &+& \left. {\left( {1 + c_1  + c_2  - c_3 } \right)\log _2 \left( {1 + c_1  + c_2  - c_3 } \right)} \right]  \notag\\
    &-&  {{1 - c} \over 2}\log _2 \left( {1 - c} \right) - {{1 + c} \over 2}\log _2 \left( {1 + c} \right) \notag\\
\end{eqnarray}

\noindent where $c \equiv \max \left\{ {\left| {c_1 } \right|,\left|
{c_2 } \right|,\left| {c_3 } \right|} \right\} $. For the mixed
state of Eq.(2), we have

\begin{eqnarray}
c_1  = {{16F^2  - 8F + 1} \over {8F^2  - 4F + 5}},c_2  =  - c_1 ,c_3
= {{12F - 3} \over {8F^2  - 4F + 5}}
\end{eqnarray}

\begin{figure}[ptb]
\includegraphics[scale=0.70,angle=0]{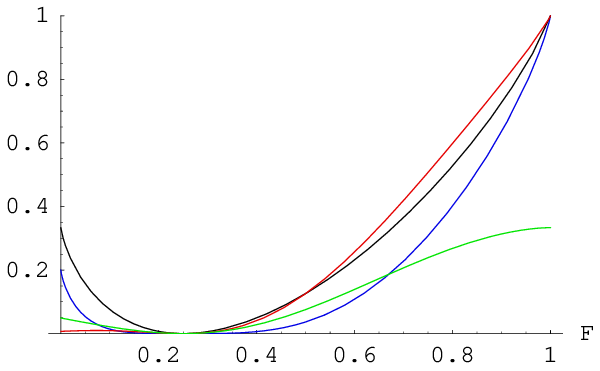}\caption{(Color online).
 The black, blue, red and green curves represent the quantum
discord as a function of $F$ correspond to $ \mathcal{D}\left( \rho
\right) $, $ \mathcal{D}\left( {\rho '} \right) $, $
\mathcal{D}\left( {\chi } \right) $ and $ \mathcal{D}\left( {\chi' }
\right) $  respectively. Here $ \rho$ denotes the initial state, $
\rho '$ denotes the intermediate state, $ \chi$ denotes the final
state, and $ \chi'$ denotes the intermediate state after performing
a random bilateral SU(2) rotation locally on each qubits.}
\label{fig1}%
\end{figure}

It is directly to see that $ c_1 $ is always a nonnegative value,
and the maximum value among them is $ \left| {c_3 } \right| $. Thus,
we can calculate the quantum discord of $ \rho '$ according to the
formula in Eq. (6). In Fig. 1 we plot the discord $
\mathcal{D}\left( {\rho '} \right) $ as a function of $F$. In order
to make a comparison with the original discord before the
purification process, we also plot $ \mathcal{D}\left( \rho  \right)
$ in Fig. 1. We can see that the quantum discord of the intermediate
state always decreases for arbitrary $F$ compared to the initial
state. Using the same method, we can also compute $
\mathcal{D}\left( \chi \right) $ and plot it in the same figure. It
is shown that quantum discord of the final state is increased after
one round of purification process, which indicates that the
purification protocol can also purify the quantum discord at the
same time. To summarize the above discussions, the quantum discord
of the original Werner states experiences two phases during one
round of purification protocol. During the first process the quantum
discord decreases when the initial Werner states are transformed
into the intermediate state. During the second process quantum
discord increases when the intermediate state is transformed into
the final state under LOCC. For the special case $ F = {1 \over 4}$,
the initial Werner state has a zero discord. In this case, it has
the form $ \rho  = \sum\nolimits_i {p_i \left| i \right\rangle
\left\langle i \right|}  \otimes \rho _i $ where $ \left| i
\right\rangle $ is the orthonormal bases. We can increase its
discord to a nonzero value by transforming this classical state into
the nonclassical form under LOCC. Thus, we conclude that LOCC can be
used to increase discord of a bipartite quantum state both for the
zero and non-zero cases.

To further understanding of the evolution of quantum discord under
LOCC. Suppose we perform a random bilateral SU(2) rotation locally
on each side of $ \rho ' $ directly and transform it into the
following Werner state: $ \chi '  = F''\left| {\beta _{11} }
\right\rangle \left\langle {\beta _{11} } \right| + {{1 - F''} \over
3}\left| {\beta _{01} } \right\rangle \left\langle {\beta _{01} }
\right| + {{1 - F''} \over 3}\left| {\beta _{10} } \right\rangle
\left\langle {\beta _{10} } \right| + {{1 - F''} \over 3}\left|
{\beta _{00} } \right\rangle \left\langle {\beta _{00} } \right|$,
with $F'' = {{2F^2  - 4F + 2} \over {8F^2  - 4F + 5}}$. We also plot
the quantum discord of $ \chi '$ in Fig. 1. We can see that the
quantum discord is decreased in this case. It indicates that we can
decrease the quantum discord by LOCC. In Fig. 2 and Fig. 3 we also
plot the dynamics of quantum mutual information and classical
correlation during the purification process, respectively. We find
that they are both increased after a round of purification protocol
for $ {1 \over 2} < F < 1 $. Under the LOCC operation from the
intermediate state $\rho '$ to the final state $\chi $ or $\chi' $,
we find that the evolution of quantum mutual information and
classical correlation is always nonincreasing.

\begin{figure}[ptb]
\includegraphics[scale=0.70,angle=0]{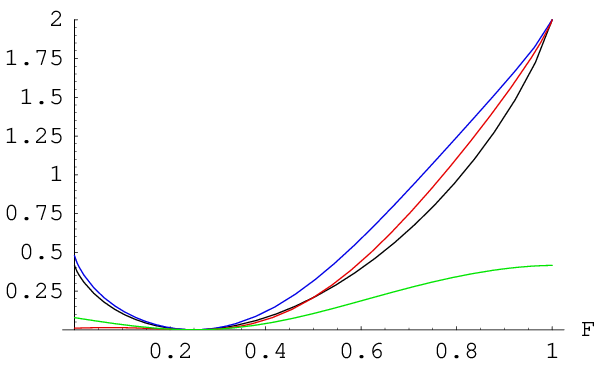}\caption{(Color online).
Dynamics of quantum mutual information $ \mathcal{I}\left( \rho
\right) $ (black line), $ \mathcal{I}\left( {\rho '} \right) $ (blue
line), $ \mathcal{I}\left( {\chi } \right) $ (red line) and $
\mathcal{I}\left( {\chi' } \right) $ (green line)  respectively.}
\label{fig1}%
\end{figure}

\begin{figure}[ptb]
\includegraphics[scale=0.70,angle=0]{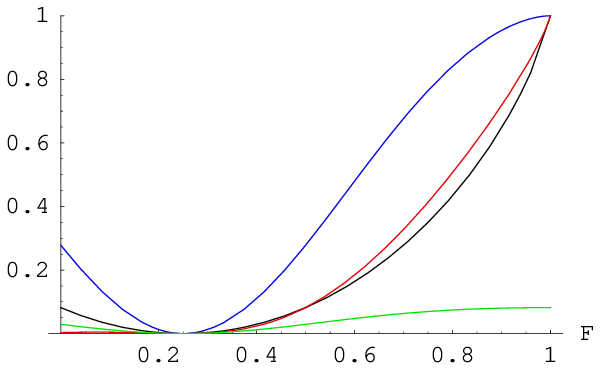}\caption{(Color online).
Dynamics of classical correlation $ \mathcal{C}\left( \rho \right) $
(black line), $ \mathcal{C}\left( {\rho '} \right) $ (blue line), $
\mathcal{C}\left( {\chi } \right) $ (red line) and $
\mathcal{C}\left( {\chi' } \right) $ (green line)  respectively.}
\label{fig1}%
\end{figure}

Before ending this paper, we want to provide a class of
higher-dimensional mixed states with computable quantum discord of
the form: $ \rho _{AB} = \sum\nolimits_{ij} {p_{ij} \left| {a_i }
\right\rangle \left| i \right\rangle \left\langle {a_j }
\right|\left\langle j \right|}$, where $ {\left| {a_i }
\right\rangle } $ is not restricted to the orthonormal bases. For
these class of state, we have $ C\left( {\rho _{AB} } \right) =
S\left( {\rho _A } \right) $ , because the minmum is obtained by
using the projector $ {\left| i \right\rangle \left\langle i
\right|} $ to get $ {\left| {a_i } \right\rangle \left\langle {a_i }
\right|} $ as the resulting state. Thus we have $ \mathcal{D}\left(
{\rho _{AB} } \right) = S\left( {\rho _B } \right) - S\left( {\rho
_{AB} } \right) $. For the general case, the classical correlations
are super-additive \cite{Henderson:2001} and thus the quantum
discord is subadditive. However, for the special mixed state $ \rho
_{AB}$\ mentioned above, we can prove its quantum discord is
additive. Recalling the Koashi-Winter monogamy relation
\cite{Koashi:2004} for quantum correlations within a pure tripartite
state $ \left| \psi \right\rangle _{ABC} $: $E_F \left( {\rho _{AC}
} \right) + C\left( {\rho _{AB} } \right) = S_A $ . Notice the
definition of discord, it is further shown that the above relation
can be rewritten as \cite{Fanchini:2011a}

\begin{eqnarray}
E_F \left( {\rho _{AC} } \right) = \mathcal{D}\left( {\rho _{AB} }
\right) + S\left( {A|B} \right) ,
\end{eqnarray}

\noindent where $ S\left( {A|B} \right) = S\left( {AB} \right) -
S\left( B \right) $ denotes the quantum conditional entropy of $\rho
_{AB}$ and $ E_F $ denote the entanglement of formation. We can
further define its regularized version \cite{Devetak:2004}
entanglement cost $ E_C \left( \rho \right) = \mathop {\lim
}\limits_{n \to \infty } {1 \over n}E_F \left( {\rho ^{ \otimes n} }
\right) $ and $ \mathcal{J}\left( {\rho _{AB} } \right) = \mathop
{\lim }\limits_{n \to \infty } {1 \over n}\mathcal{D}\left( {\rho ^{
\otimes n} } \right)$, then we have:

\begin{eqnarray}
E_C\left( {\rho _{AC} } \right) = \mathcal{J} \left( {\rho _{AB} }
\right) + S\left( {A|B} \right)
\end{eqnarray}

It is shown in Ref.\cite{Cornelio:2011} that the equality above
gives a operational meaning to the regularized version of quantum
discord as a measure of the amount of entanglement loss when Alice
and Bob distill entanglement by hashing. If we choose $ {\rho _{AB}
} $ as the above mixed states, then we can write its purified form
as: $ \left| \psi \right\rangle _{ABC} = \sum\limits_i {\lambda _i
\left| {a_i } \right\rangle \left| i \right\rangle \left| {b_i }
\right\rangle } $, where $ {\left| {b_i } \right\rangle } $ may not
be orthogonal for different $i$. It is directly to see that $ {\rho
_{AC} } $ is a separable state with zero entanglement formation and
entanglement cost. Thus by Eq.(8) and Eq.(9) we have that quantum
discord of the special mixed state $\rho _{AB}$ mentioned above is
additive.

In the above discussions we have investigated the dynamics of
quantum discord during the purification process. It is found that
quantum discord is increased after a round of purification protocol.
We show that during the original BBPSSW purification protocol
quantum discord can also be increased at the same time. Therefore,
we conclude that quantum discord of arbitrary two-qubit state can be
increased by the original purification protocol. However, for the
higher-dimensional case, we do not know whether quantum discord is
also increased. At present, we cannot answer this question due to
the lack of analytical formula of quantum discord for the
higher-dimensional case.

Wei Song was supported by National Natural Science Foundation of
China under Grant No.10905024, the Key Project of Chinese Ministry
of Education under Grant No.211080, the Doctoral Startup Foundation
of Hefei Normal University under Grant No.2011rcjj03 and the Key
Program of the Education Department of Anhui Province under Grant
No. KJ2011A243. The Centre for Quantum Technologies is funded by the
Singapore Ministry of Education and the National Research Foundation
as part of the Research Centres of Excellence program. Ming Yang was
supported by the Key Project of Chinese Ministry of
Education.(No.210092). Zhuo-Liang Cao was supported by the National
Natural Science Foundation of China under Grant No.11005029.


\begin{thebibliography}{99}


\bibitem {Neilsen:2000}M. A. Neilsen and I. L. Chuang, \textit{Quantum
Computation and Quantum Information} (Cambridge University Press,
New York, 2000).

\bibitem {Bennett:1999}C. H. Bennett, D. P. DiVincenzo, C. A. Fuchs, T. Mor, E.
Rains, P. W. Shor, J. A. Smolin, and W. K. Wootters, Phys. Rev. A
59, 1070 (1999).

\bibitem {Horodecki:2005}M. Horodecki, P. Horodecki, R. Horodecki, J. Oppenheim, A.
Sen, U. Sen, and B. Synak-Radtke, Phys. Rev. A 71, 062307 (2005).

\bibitem {Braunstein:1999}S. L. Braunstein, C. M. Caves, R. Jozsa, N. Linden, S. Popescu,
and R. Schack, Phys. Rev. Lett. 83, 1054 (1999).

\bibitem {Meyer:2000}D. A. Meyer, Phys. Rev. Lett. 85, 2014 (2000).

\bibitem {Datta:2005}A. Datta, S. T. Flammia, and C. M. Caves, Phys. Rev. A 72,
042316 (2005); A. Datta and G. Vidal, \emph{ibid.} 75, 042310
(2007).

\bibitem {Henderson:2001}L. Henderson and V. Vedral, J. Phys. A 34, 6899 (2001).

\bibitem {Zurek:2000}W. H. Zurek, Annalen der Physik, 9, 855 (2000).

\bibitem {Ollivier:2001}H. Ollivier and W. H. Zurek, Phys. Rev. Lett. 88, 017901 (2001).


\bibitem {Rajagopal:2002}A. K. Rajagopal and R. W. Rendell, Phys. Rev. A 66,
022104 (2002).

\bibitem {Luo:2008}S. Luo, Phys. Rev. A 77, 022301 (2008).

\bibitem {Modi:2010}K. Modi, T. Paterek, W. Son, V. Vedral, and M. Williamson, Phys. Rev. Lett. 104, 080501
(2010).

\bibitem {Dakic:2010}B. Dakic, V. Vedral, and C. Brukner, Phys. Rev. Lett. 105, 190502 (2010).


\bibitem {Chen:2010}L. Chen, E. Chitambar, K. Modi, and G. Vacanti, Phys. Rev. A 83, 020101 (R)
(2011).


\bibitem {Ferraro:2010} A. Ferraro, L. Aolita, D. Cavalcanti, F. M. Cucchietti, A.
Acin, Phys. Rev. A 81, 052318 (2010).

\bibitem {Knill:1998}E. Knill and R. Laflamme, Phys. Rev. Lett. 81, 5672
(1998).

\bibitem {Datta:2008}A. Datta, A. Shaji, and C. M. Caves, Phys. Rev.
Lett. 100, 050502 (2008).

\bibitem {Cornelio:2011}M. F. Cornelio, M. C. de Oliveira and F. F. Fanchini,
Phys. Rev. Lett. 107, 020502 (2011).

\bibitem {Fanchini:2011}F. F. Fanchini, M. C. de Oliveira, L. K. Castelano, M. F.
Cornelio, arxiv:1110.1054 (2011).

\bibitem {Piani:2011}M. Piani, S. Gharibian, G. Adesso, J. Calsamiglia, P. Horodecki and
A. Winter, Phys. Rev. Lett. 106, 220403 (2011).

\bibitem {Streltsov:2011}A. Streltsov, H. Kampermann and D. Bruss, Phys. Rev. Lett. 106,
160401 (2011).

\bibitem {Shabani:2009}A. Shabani and D. A. Lidar, Phys. Rev. Lett. 102,
100402 (2009).

\bibitem {Cavalcanti:2011}D. Cavalcanti, L. Aolita, S. Boixo, K. Modi, M. Piani and A. Winter,
Phys. Rev. A 83, 032324 (2011).

\bibitem {Madhok:2011}V. Madhok and A. Datta, Phys. Rev. A 83, 032323 (2011).


\bibitem {Fanchini:2011d}F. F. Fanchini, L. K. Castelano, M. F. Cornelio, M. C. de Oliveira,
 New Journal of Physics 14, 013027 (2012).

\bibitem {Werlang:2009}T. Werlang, S. Souza, F. F. Fanchini, and C. J. Villas Boas, Phys. Rev. A 80, 024103 (2009).


\bibitem {Maziero:2009}J. Maziero, L. C. Celeri, R. M. Serra, and V. Vedral, Phys. Rev. A 80, 044102 (2009).

\bibitem {Yu:2009}T. Yu and J. H. Eberly, Science. 323, 598 (2009).

\bibitem {Wang:2010}B. Wang, Z.-Y. Xu, Z.-Q. Chen, and M. Feng, Phys. Rev.
A 81, 014101 (2010).

\bibitem {Fanchini:2010} F. F. Fanchini, T. Werlang, C. A. Brasil, L. G. E. Arruda, A. O.
Caldeira, Phys. Rev. A 81, 052107 (2010).

\bibitem {Xu:2010a}Jin-Shi Xu, Xiao-Ye Xu, Chuan-Feng Li, Cheng-Jie Zhang, Xu-Bo Zou, and Guang-Can Guo, Nat. Commun. 1, 7 (2010).


\bibitem {Xu:2010b}Jin-Shi Xu, Chuan-Feng Li, Cheng-Jie Zhang,
Xiao-Ye Xu, Yong-Sheng Zhang, and Guang-Can Guo, Phys. Rev. A 82,
042328 (2010).


\bibitem {Horodecki:1998}M. Horodecki, P. Horodecki, and R. Horodecki, Phys. Rev.
Lett. 80, 5239 (1998).


\bibitem {Bennett:1996}C. H. Bennett, G. Brassard, S. Popescu, B. Schumacher, J. A.
Smoin, and W. K. Wootters, Phys. Rev. Lett. 76, 722 (1996).

\bibitem {Werner:1989}R. F. Werner, Phys. Rev. A 40, 4277 (1989).

\bibitem {Hamieh:2004}S. Hamieh, R. Kobes, and H. Zaraket, Phys. Rev. A 70, 052325 (2004).

\bibitem {Lang:2010}M. D. Lang, and C. M. Caves, Phys. Rev. Lett. 105, 150501 (2010).


\bibitem {Luo:2008b}S. Luo, Phys. Rev. A 77, 042303
(2008).

\bibitem {Koashi:2004}M. Koashi and A. Winter, Phys. Rev. A 69, 022309 (2004).



\bibitem {Fanchini:2011a} F. F. Fanchini, M. F. Cornelio, M. C. de Oliveira, A. O.
Caldeira, Phys. Rev. A 84, 012313 (2011).

\bibitem {Devetak:2004}I. Devetak and A. Winter, IEEE Tran. Info. Theory. 50, 3183 (2004).

\end{thebibliography}
\end{document}